\documentclass[
superscriptaddress,
amsmath,amssymb,
 reprint,
 jmp,
 prb,
 floatfix,
]{revtex4-1}
\usepackage{comment}
\usepackage{graphicx}\usepackage{dcolumn}\usepackage{bm}\usepackage{hyperref}
\usepackage[dvipsnames]{xcolor}
\usepackage{gensymb}
\usepackage{pdfcomment}

\hypersetup{colorlinks=true,
    linkcolor=blue,
    citecolor=blue,
    urlcolor=cyan,
}

\begin{document}

\preprint{AIP/123-QED}

\title[Manuscript in preparation for Phys. Rev. B.]{On the origin of magnetization auto-oscillations in constriction-based spin Hall nano-oscillators}
\author{Mykola Dvornik}
\author{Ahmad A. Awad}

\affiliation{Department of Physics, University of Gothenburg, 412 96 Gothenburg, Sweden}
\author{Johan \AA kerman}
\affiliation{Department of Physics, University of Gothenburg, 412 96 Gothenburg, Sweden}
\affiliation{Materials and Nanophysics, School of ICT, KTH Royal Institute of Technology, 164 00 Kista, Sweden}

\date{\today}             
\begin{abstract}
We use micromagnetic simulations to map out and compare, the linear and auto-oscillating modes in constriction-based spin Hall nano-oscillators as a function of applied magnetic field with varying magnitude and out-of-plane angle. We demonstrate that for all possible applied field configurations the auto-oscillations emerge from the localized linear modes of the constriction. For field directions tending towards the plane, these modes are of the so-called "edge" type, \emph{i.e.}~localized at the opposite sides of the constriction. When the magnetization direction instead approaches the film normal, the modes transform to the so-called "bulk" type, \emph{i.e.}~localized \emph{inside} the constriction with substantially increased precession volume, consistent with the re-distribution of the magnetic charges from the sides to the top and bottom surfaces of the constriction. In general, the threshold current of the corresponding auto-oscillations increases with the applied field strength and decreases with its out-of-plane angle, consistent with the behavior of the internal field and in good agreement with a macrospin model. A quantitative agreement is then achieved by taking into account the strongly non-uniform character of the system via a mean-field approximation. Both the Oe field and the spin transfer torque from the drive current increase the localization and decrease the frequency of the observed mode. Furthermore, the anti-symmetric Oe field breaks the lateral symmetry, favoring the localized mode at one of the two constriction edges, in particular for large out-of-plane field angles where the threshold current is significantly increased and the edge demagnetization is suppressed.

\end{abstract}

\pacs{Valid PACS appear here}                             

\maketitle

\section{Introduction}
It is well known that ferromagnetic insulators can be excited into strongly nonlinear magnetodynamical states by the application of sufficiently strong rf magnetic fields \cite{DeGasperis1987,Chen1993,Bauer1998, Buettner2000}. The same approach, although possible, is rather inefficient for magnetic metals, as they typically exhibit much higher magnetic losses\cite{Demidov2010,Keatley2011}. However, with the emergence of spin transfer torque\cite{slonczewski1996jmmm,berger1996prb,ralph2008jmmm} (STT) it has become possible to excite and sustain highly non-linear, nano-scale magnetization dynamics in metals, including propagating spin waves,\cite{tsoi1998prl,slonczewski1999jmmm, madami2011nn,madami2015prb} localized bullets,\cite{rippard2004prl,slavin2005prl,bonetti2010prl} vortices,\cite{mistral2006apl} and droplets\cite{mohseni2013sc,macia2014ntn,chung2014jap,chung2015ltp,chung2016ntc}. 
The majority of these studies have been devoted to extended geometries where dissipative magnetic solitons are typically nucleated by employing the negative nonlinearity\cite{slavin2005ieeem} of the system that pushes the original FMR mode into the fundamental magnonic bandgap, where propagation of spin waves is ultimately forbidden. This results in self-localization of the magnetization dynamics in the vicinity of the spin-polarized current source. However, patterned magnetic structures support natural confinement of the magnetization dynamics, in the form of so-called  “edge” magnonic modes\cite{Jorzick2002}. These excitations are again typically observed in the fundamental bandgap, similar to the dissipative magnetic solitons.  Since nano-patterned materials are at the core of the emerging spintronics-based technologies,  it is essential to understand their response to the application of spin-polarized currents.
	
Prominent examples of such systems are the so-called nano-constriction\cite{demidov2014apl, Mazraati2016} and nanowire\cite{duan2014ntc, Yang2015srp} based spin Hall nano-oscillators\cite{chen2015pieee,demidov2012ntm,ranjbar2014ieeeml} (SHNOs), where pure spin currents are injected from the heavy metal (such as Pt or W) to the adjoint ferromagnetic layer (e.g. NiFe or CoFeB). In constriction and wires with widths below 200 nm, the injected spin current density is sufficient to nucleate self-sustained magnetization dynamics and then drive it into a strongly nonlinear regime. In contrast to extended geometries, however, the auto-oscillations in nanowire SHNOs have been shown to emerge from the semi-confined linear modes of the bulk and edge types\cite{duan2014ntc}. It was later demonstrated, that the edge mode becomes further localized with the increase of its amplitude\cite{Yang2015srp}. Due to the significant shrinking of the nonlinear edge mode it shows a much-reduced linewidth, as it interacts and scatters  less with other modes. This property is essential for successful application of SHNOs for microwave signal generation.

Although constriction-based SHNOs show even lower linewidths and, in addition, an unprecedented ability to establish mutual synchronization over large distances and device counts in  out-of-plane fields\cite{Awad2016}, their dynamics has not yet been analyzed in detail. Kendziorczyk and Kuhn\cite{Kendziorczyk2016prb} simulated current-driven dynamics of constriction-based SHNOs in in-plane fields and demonstrated that the auto-oscillations are strongly localized to the constriction edges, consistent with the appearance of minima in the static internal field. While this might suggest that auto-oscillations originate from the linear localized mode of the nanoconstriction, the relation between the two was not investigated. To contrast, our micromagnetic simulations of mutually synchronized SHNOs in close to perpendicular fields demonstrated that mutual synchronization was possible to establish thanks to auto-oscillation modes \emph{inside} the constriction and further extending into the SHNO leads.\cite{Awad2016} A detailed study of the out-of-plane angular dependence of both linear and auto-oscillating spin wave modes in constriction-based SHNOs is hence required, both to establish their relation, and investigate a possible cross-over from edge to bulk localization. 
Here we employ micromagnetic simulations to demonstrate the origin and spatial properties of the auto-oscillations in constriction-based SHNOs for a wide range of field magnitudes and out-of-plane angles. We show that the field dependence of the threshold current agrees quite well with a macrospin model that neglects SW radiation losses; the agreement becomes essentially perfect when the macrospin model is refined using a mean-field approach. We then explicitly demonstrate that for all field angles, the auto-oscillations emerge from the localized linear modes of the constriction. For fields tending towards the plane, these localized modes reside at the constriction edges. However, when the magnetization direction approaches the film normal, the edge modes move into the interior of the constriction, transforming into a so-called "bulk" type with significantly increased volume of precession. 

\section{Micromagnetic simulations}

\begin{figure}\centering
\includegraphics[width=8.6cm]{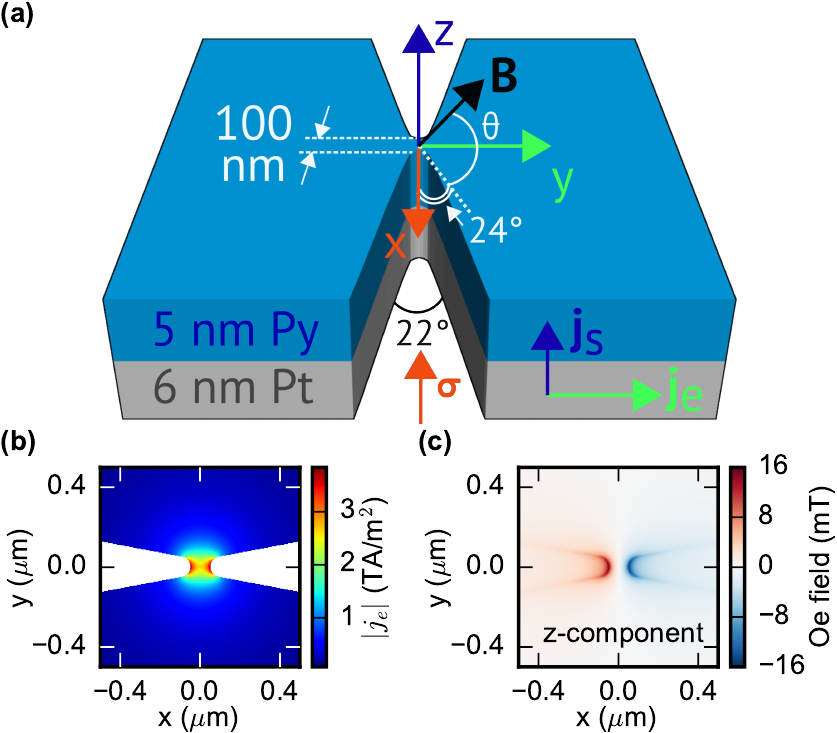}
\caption{(a) Schematic of the simulated Py/Pt constriction SHNO. (b) Magnitude of the lateral electrical current density in Pt, and (b) out-of-plane component of the Oersted field in Py calculated for an applied current of 2 mA.
}
\label{fig:1}
\end{figure}

Here we simulate a stack of 6 nm Pt and 5 nm Py layers containing a round-shaped nano-constriction of 100 nm width, opening angle of 22$^{\degree}$ and curvature radii of 50 nm as schematically shown in Fig.~\ref{fig:1}(a). The particular choice of sample geometry follows those already investigated experimentally in the literature. The electrical current density and the corresponding Oe field were simulated in COMSOL \cite{COMSOL} assuming the full-scale Pt/Py bilayer and an electrical current of I$_{ref}$ = 2 mA, while linear scaling is assumed for any other values. The Oe field and current density were sampled at the Py and Pt sites, respectively. The data is then estimated on the rectangular mesh that matches to the micromagnetically simulated domain. The corresponding profiles are shown in Figs.~\ref{fig:1}(a) and (b). We assumed that the electrical current density in Pt, $\mathbf{j}_{e}$, leads to a pure spin current injection into Py along the interface normal with the magnitude of $j_{s} = (I/I_{ref}) \theta_{SH} \left| \mathbf{j}_{e} \right|$, where $\theta_{SH}$ = 0.08 is the Pt spin Hall angle and $I$ is the applied current. Although the applied current bends in the vicinity of the constriction edges, it is still dominated by the longitudinal component (y-axis). So we assume that, in accordance with the properties of the spin Hall effect, the injected spin current is uniformly polarized anti-parallel to the x-axis, i.e. $\mathbf{\sigma}=-\mathbf{x}$.

The micromagnetic simulations are carried out using the mumax3 solver\cite{vansteenkiste2014aip} with the input provided by the COMSOL simulations described above. Although the structure includes a heavy metal layer, only the ferromagnetic part is explicitly considered in the simulations. The corresponding Py layer has dimensions of 2 um $\times$ 2 um $\times$ 5 nm subdivided into a rectangular mesh of $\Delta x \times \Delta y \times \Delta z$ = 3.9 $\times$ 3.9 $\times$ 5 nm$^3$ cells. Due to the difference in electrical resistances of Pt and Py layers, the current mostly flows through the heavy metal. So any contribution of the current going via the ferromagnet to the magnetization dynamics, \emph{e.g.} via a (non-adiabatic) spin transfer torque, is therefore neglected in the micromagnetic simulations. The Py/Pt stack is assumed to have a saturation magnetization of $\mu_0 M_s$ = 0.754 T, a Gilbert damping of 0.02, a gyromagnetic ratio of 29.53 GHz/T and an exchange stiffness of 10 pJ/m, consistent with our experimental studies. \cite{Awad2016,Durrenfeld2017}

The magnetic field, \textbf{B$_{0}$}, is applied at a fixed in-plane angle of 24$^{\degree}$ as it maximizes the AMR read-out from the SHNO.\cite{Awad2016} The applied field strength and out-of-plane angle are then varied.  The magnetization dynamics is simulated by integrating the Landau-Lifshits-Gilbert-Slonczewski equation over 187 ns, with the first 62 ns discarded in the subsequent analysis to exclude transient effects. For the sake of consistency, the linear eigenmodes of the system are estimated at the threshold current by taking into account both the Oe field and the STT. To avoid auto-oscillations, the damping of the system is increased 1.02 times and then the system is excited by the sinc rf field with the amplitude of 1 mT and a cut-off frequency of 40 GHz. The linear response is captured over 125 ns. The frequencies and spatial profiles of linear and auto-oscillating modes of the system are extracted using methods explained elsewhere\cite{dvornik2011numerical,Dvornik2013}. 

\section{Results and discussion}

For any given configuration of the applied magnetic field, we first want to estimate the auto-oscillation threshold current, $I_{th}$. For this purpose, at a given value of the applied current, we first run the simulations for 5 ns, where the system undergoes some transient behavior, and over the next 10 ns monitor the behavior of the maximum torque; if the maximum torque increases, we assume that auto-oscillations have started. Using this criterion, we then employ the so-called "bisect" method to estimate the threshold current in a range of [0.5, 5.0] mA by iteratively shrinking this interval until its bounds are separated by 1 $\mu$A, which gives $I_{th}$ $\pm$ 0.5 $\mu$A. Compared to the total energy, the torque shows a significantly smaller degree of numerical noise \cite{vansteenkiste2014aip}, which makes it more suitable for this particular type of analysis.

\begin{figure}\centering
\includegraphics[width=8.6cm]{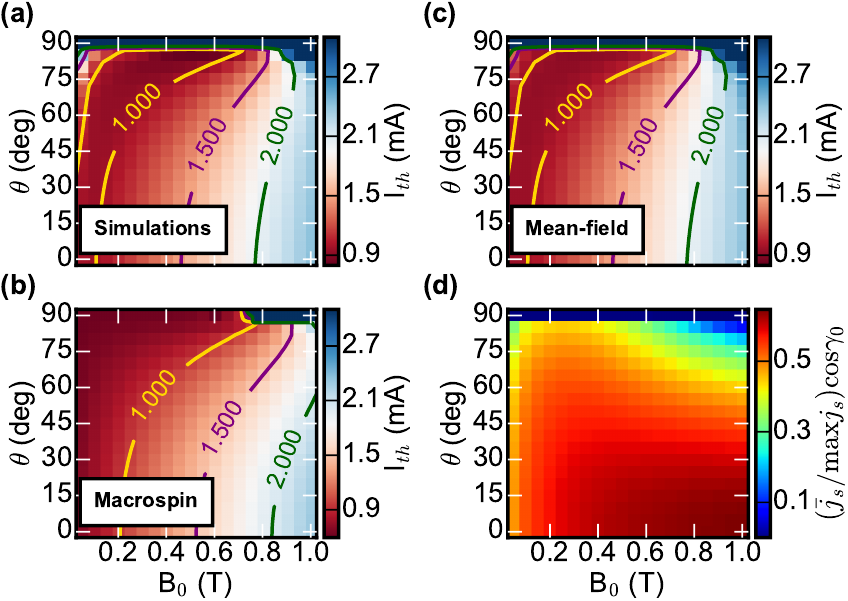}
\caption{Auto-oscillation threshold current vs.~applied field strength and out-of-plane angle as estimated (a) using micromagnetic simulations, (b) a macrospin model given by Eqs. (\ref{ith}) and (\ref{bint}), and (c) mean-field model given by Eqs. (\ref{macrospin}) and (\ref{meanfield}). (d) The efficiency of pure spin current injection versus applied field geometry. }
\label{fig:2}
\end{figure}

The results of these calculations are shown in Fig.~\ref{fig:2}(a). In general, we find that $I_{th}$ increases with field strength and decreases with increasing field angle. This behavior can be understood using the model developed in Ref. \cite{slavin2005ieeem}: assuming a macrospin approximation (no propagating SWs), isotropic spin polarization efficiency $\epsilon = 1$,  easy plane shape anisotropy, and neglecting any magneto-crystalline anisotropy, the following expression holds: \begin{equation}
    I_{th} = \frac{\alpha}{\sigma_{0} \cos \gamma_{0}} \left( \omega_{B} + \frac{\omega_{M}}{2} \right),
\label{ith}
\end{equation}
where $\omega_{M} = \gamma \mu_{0} M_{s}  \cos \theta$, $\sigma_{0} = \frac{ \epsilon \gamma \hbar}{2eM_{s}St}$ is the magnitude of the STT, $S = I_{ref} / \max j_{s}$ is the effective area of pure spin current injection, $\gamma_{0} = \theta - \theta_{p}$ is the angle between the equilibrium magnetization and the polarization of the pure spin current (chosen to achieve anti-damping for the positive applied currents), and $\theta_{p}$ and  $\theta$ are the the out-of-plane angles of the spin current polarization 
and the equilibrium magnetization, respectively. $\omega_{B} = \gamma B$, where $B$ is the magnitude of the internal magnetic field, which, according to the magneto-static boundary conditions, reads
\begin{equation}
B = B_{0} \sqrt{1 + \frac{\mu_{0} M_{s}}{B_{0}}\sin \theta (\frac{\mu_{0} M_{s}}{B_{0}} \sin \theta - 2 \sin \theta_{0})}. 
\label{bint}
\end{equation}
It follows from Eqs. (\ref{ith}) and (\ref{bint}) that the threshold current increases with the magnitude of the applied field and decreases with its out-of-plane angle, consistent with our simulations. 

However, at high fields and large angles, which make the magnetization approach the film normal, the threshold current again increases due to (a) an increase of the internal field given by Eq. (\ref{bint}), and (b) a decrease of the STT efficiency as $\gamma_{0}$ approaches $\pi/2$. For applied fields close to or exceeding $\mu_{0} M_{s}$, the out-of-plane angles of the equilibrium magnetisation and the applied field increase simultaneously, so that both contributions to $I_{th}$ counteract each other. This explains the observed flattening, or even increase, of the threshold current at high fields, and is captured well by both simulations (Fig.~\ref{fig:2}(a)) and the macrospin model (Fig.~\ref{fig:2}(b)), which agree, at least qualitatively, in this region.

The agreement is, however, substantially worse below 0.7 T, where an increase in $I_{th}$ is observed in the simulations despite an increase of the out-of-plane angle of the applied field and virtually no changes to the direction of the equilibrium magnetization. This could be attributed to the reduction of the STT efficiency, \emph{e.g.} due to the rotation of the magnetization in-plane, which would reduce $\gamma_{0}$. To further examine the validity of the model given by Eq. (\ref{ith}), we employ a mean-field approach to estimate the relevant parameters, given by the set $\wp=\{B,\ \theta, \gamma_{0}, j_{s} \}$, from the simulations as follows:

\begin{equation}
    \bar{\wp} = \frac{\sum_{i} \sum_{j} \wp_{ij} m_{ij}^2}{\sum_{i} \sum_{j} m_{ij}^2}
    \label{meanfield}
\end{equation}
where $m_{ij}$ is the spatial profile of the auto-oscillations amplitude and the bar symbol denotes the averaged value. The summation is performed over the 512 $\times$ 512 nm$^2$ domain around the nano-constriction, where most of the auto-oscillation amplitude is localized. Since the spin current is strongly non-uniform, and assuming that the spatial profiles of the auto-oscillations can change with the applied field, the STT magnitude is expected to be mode-specific. To account for this effect, we re-normalize the STT magnitude to $\sigma_{0}^{\prime} = \sigma_{0} \bar{j}_{s} /\max j_{s}$, and finally get
\begin{equation}
I_{th} = \frac{\gamma \alpha}{\sigma_{0}^{\prime} \cos \bar{\gamma}_{0}} \left(\bar{B} + \frac{1}{2}\mu_0 M_s \cos \bar{\theta} \right)
\label{macrospin}
\end{equation}
The data calculated using Eq. (\ref{macrospin}) is shown in Fig.~\ref{fig:2}(c). A quite remarkable \emph{quantitative} agreement with the simulation is observed. The agreement is achieved without including any radiation losses (\emph{i.e.}~propagating SWs) thus confirming the localized character of the auto-oscillations, consistent with Ref. \cite{demidov2014apl, Kendziorczyk2016prb}. The increasing $I_{th}$ at high out-of-plane angle and weak applied fields is now fully recovered, consistent with the reduction of the pure spin current injection efficiency as shown in Fig.~\ref{fig:2}(d). In particular, we not only observe the in-plane rotation of the equilibrium magnetization, as captured by the $\cos \gamma_{0}$ term, but also the reduction of the mean-field value of the injected pure spin current magnitude, i.e. $\bar{j}_{s} / \max j_{s}$, which confirms changes of the auto-oscillations vs.~applied field geometry.

\begin{figure}\centering
\includegraphics[width=8.6cm]{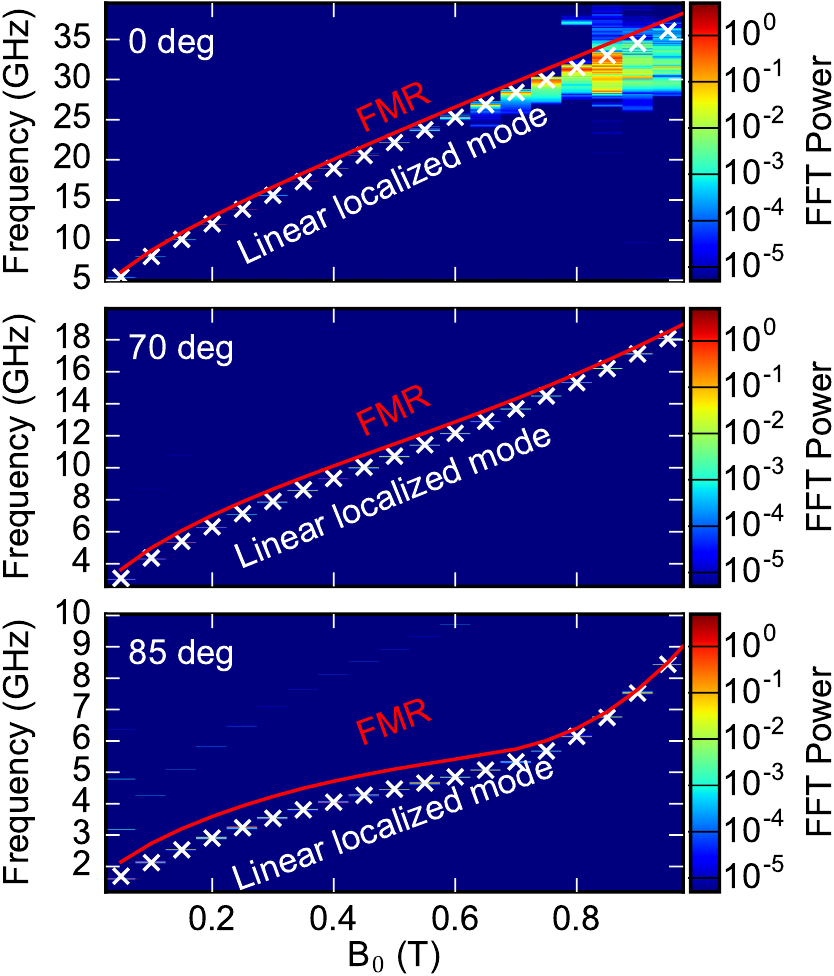}
\caption{Auto-oscillation spectral density vs.~applied field strength and three different field angles. Red lines and white cross signs show frequencies of the FMR and linear localized modes, respectively. 
}
\label{fig:3}
\end{figure}

The auto-oscillation power spectra and linear eigenmodes of the nanoconstriction, calculated at the threshold current for various field geometries, are shown in Fig.~\ref{fig:3}. The auto-oscillations always appear below the frequency of the (quasi) uniform ferromagnetic resonance (FMR), which again confirms their localized character. However, in contrast to the extended geometries, where non-linearity driven self-localization of the SWs happens, in our present case, the auto-oscillations essentially coincide with the linear localized modes of the nanoconstriction, similar to the nanowire SHNOs. Although our system should also support regular SW bullets, they are not observed in the simulations, since the FMR amplitude is negligible inside the nanoconstriction, where most of the spin current injection happens.

\begin{figure}[t]
\centering
\includegraphics[width=8.6cm]{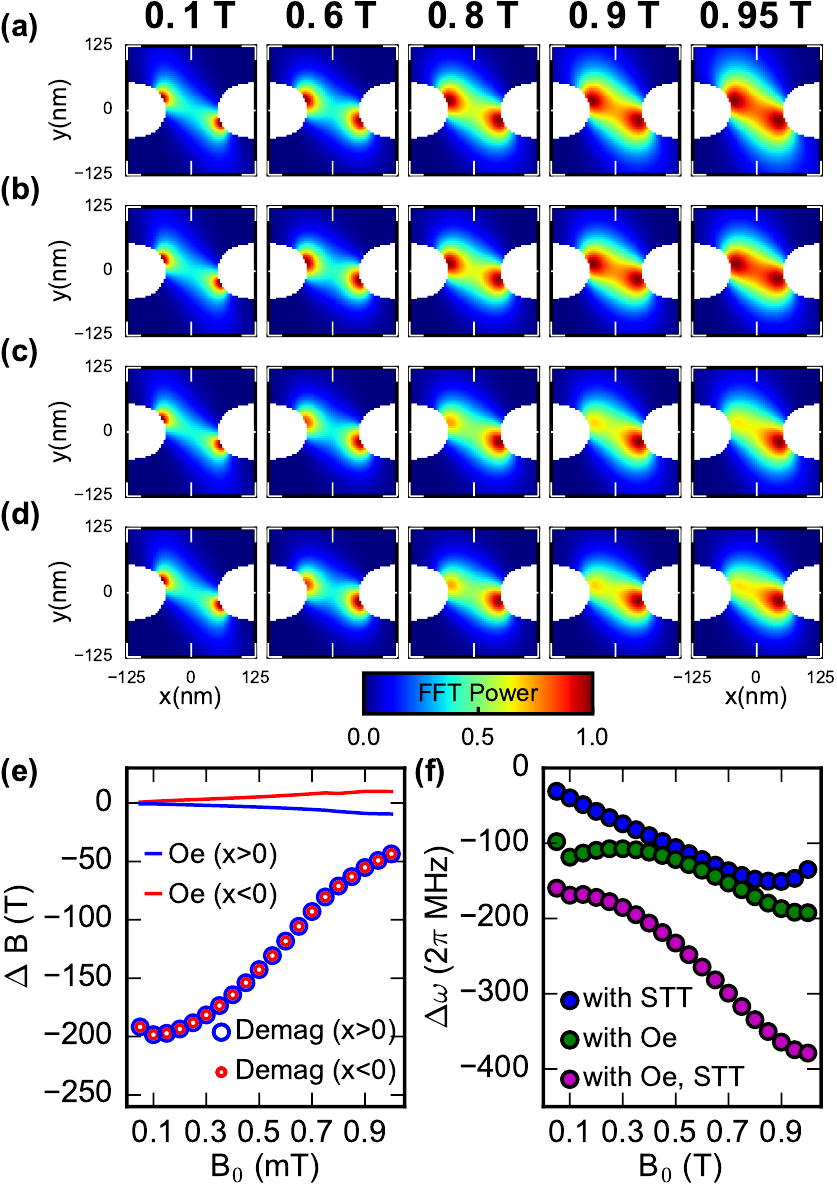} 
\caption{Spatial profiles of the auto-oscillations calculated for a field applied at $\theta$ = 70$\degree$ and (a) without any Oe field or STT, (b) with STT, (c) with Oe field, and (d) with both Oe field and STT. (e) Contribution of the demagnetizing field and Oe field to the depth of the spin-wave wells vs.~applied field strength. (f) Contribution of the STT and Oe field to the frequency of the edge mode vs.~applied field strength.}
\label{fig:4}
\end{figure}

To investigate the fundamental origin of the localization we calculate the spatial profiles of the linear modes simulated at five different applied fields and using four different combinations of having the Oe field and the STT terms on/off during the simulation: Fig.~\ref{fig:4}(a) includes neither the Oe field nor STT, Fig.~\ref{fig:4}(b) includes only STT, Fig.~\ref{fig:4}(c) includes only the Oe field, and Fig.~\ref{fig:4}(d) includes both. Comparing Fig.~\ref{fig:4}(a) with Fig.~\ref{fig:4}(b)-d, it is first of all clear that neither the Oe field nor STT are required for confinement. The confinement can hence be explained by the demagnetization field alone and the observed dynamics is essentially an edge mode typically observed in patterned magnetic structures\cite{Jorzick2002} where the demagnetizing field in the vicinity of the edges creates local minima in the effective magnetic field, \emph{i.e.~}so-called "SW wells". This field is produced by the surface magnetic charges that emerge due to the divergence of the normal-to-surface component of the equilibrium magnetization\cite{Joseph1965, Bryant1989}. In our particular geometry, the magnetic charges on the opposite edges of the constriction arise from the divergence of the in-plane component of the equilibrium magnetization. 

\begin{figure*}\centering
\includegraphics[width=\textwidth]{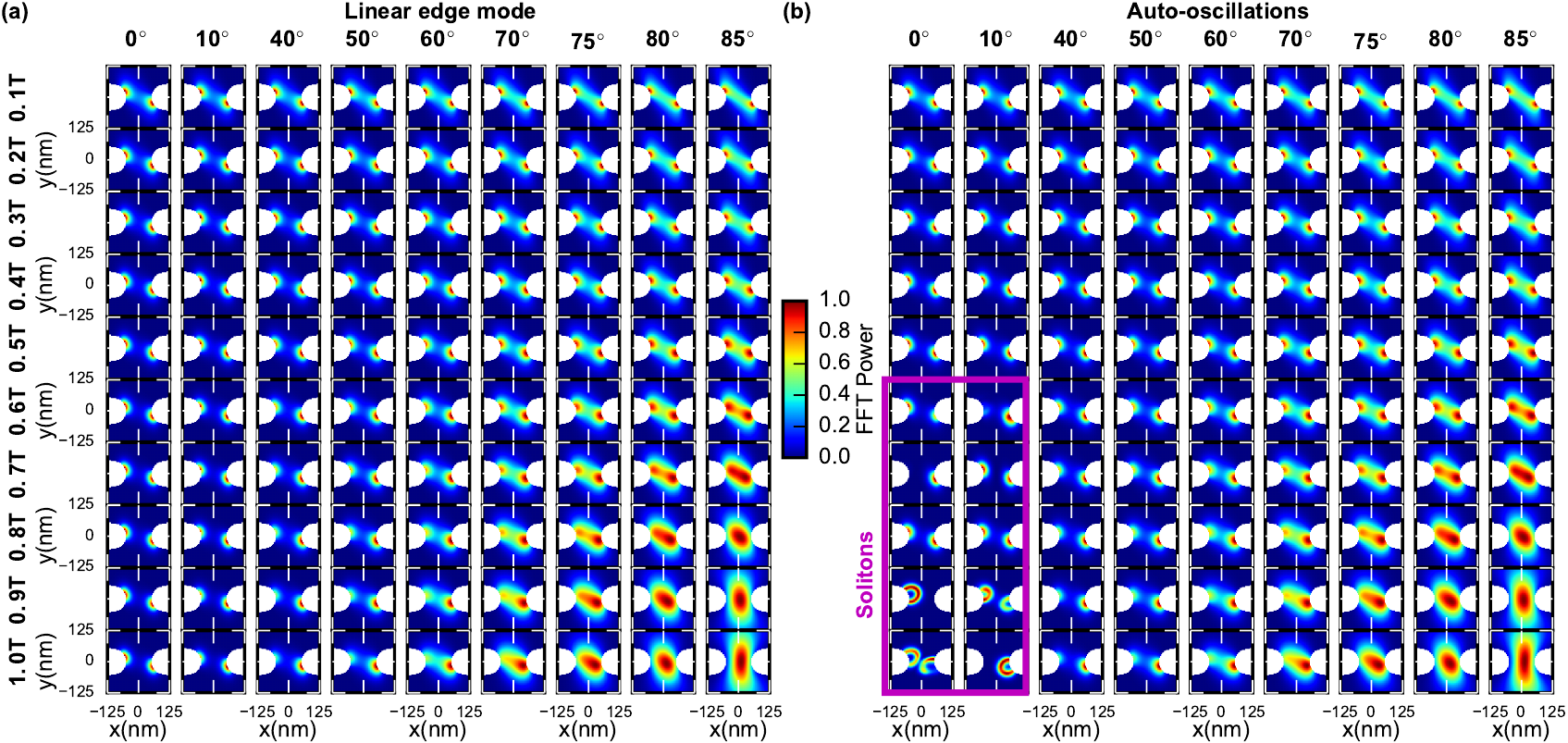}
\caption{Spatial profiles of (a) the linear, and (b) the auto-oscillating, modes, simulated at unit supercriticality for applied fields with different strengths and out-of-plane angles. }
\label{fig:5}
\end{figure*}

When we include STT, a slight reduction of the mode area is observed in strong fields (Fig.~\ref{fig:4}(b)), consistent with how the spin current affects the internal field. In our simulations STT always counteracts the applied field and has its strongest contributions at the edges where the current density is the highest (Fig.~\ref{fig:1}(b)). STT hence increases the depth of the SW wells and therefore enhances the localization of the edge modes.

The inclusion of the Oe field leads to a significant asymmetry (Fig.~\ref{fig:4}(c)) of the (otherwise symmetric) edge modes, again primarily in strong out-of-plane fields. This is a direct consequence of the asymmetric nature of the out-of-plane component of the Oe field, with respect to the constriction center (see Fig.~\ref{fig:1}(c)). The Oe field hence suppresses the SW well on one side of the constriction, while it strongly enhances the mode localization on the other. As both the Oe and STT contributions are proportional to the applied current, which is varied in our simulations to stay at the onset of the auto-oscillations, they are stronger in oblique fields where the threshold current is larger.

\begin{figure}\centering
\includegraphics[width=8.6cm]{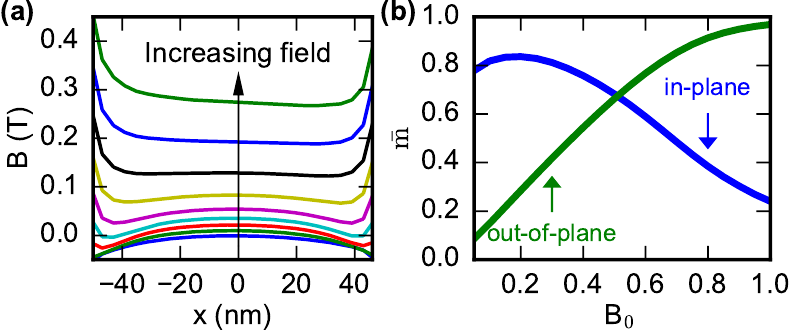} 
\caption{(a) Effective magnetic field sampled along the constriction width vs.~fields applied at $\theta=$ 85$\degree$ ranging from 0.2 T to 1 T in steps of 0.1 T. (b) Mean-field values of the in-plane and out-of-plane components of the equilibrium magnetization.}
\label{fig:6}
\end{figure}

To quantify both effects, we first estimate the contributions of the demagnetizing and the Oe fields to the internal field using the magnetostatic boundary conditions. In particular, we calculate their projections on the equilibrium magnetization. If the corresponding projection is positive (negative), then it adds up to (subtracts from) the internal field, \emph{i.e.}~suppresses (enhances) the SW wells. Finally, we calculate the contribution of both fields to the depth of the SW wells on the opposite sides of the constriction (i.e.~$x > 0$ and $x < 0$) as $\Delta B^{i} = B^{i}(\mathbf{min}) - B^{i}(\mathbf{0})$, where $\mathbf{0}$ is the coordinate of the constriction center, $i$ denotes the corresponding field contribution, and $\mathbf{min}$ is the coordinate of the minimum in projection of the demagnetizing field.  The result is shown Fig.~\ref{fig:4}(d). We note that for weak and moderate applied fields the contribution of the Oe field is negligible compared to the demagnetization. However, at higher fields, the combined effect of \emph{i}) a weakening demagnetization, due to a decreasing in-plane component of the equilibrium magnetization as it tilts out-of-plane, and \emph{ii}) an increasing threshold current, rapidly increases the role of the Oe field. In Fig.~\ref{fig:4}(f) we finally plot how much the Oe field and STT shift the frequency of the edge mode and conclude that their contributions are comparable. 
We now turn to the spatial profiles of the linear modes across a more complete range of field angles and strengths (Fig.~\ref{fig:5}(a)). At most conditions, the edge mode clearly dominates and, as discussed above, remains mostly symmetric at low to moderate field strengths. At high fields and intermediate field angles, the antisymmetric influence of the Oe field is clearly visible.  However, as the strength and the out-of-plane angle of the applied field increase further (\emph{i.e.} towards the bottom right corner of the figure), we observe a fundamental change of the spatial profile: the edge mode first localizes further, then de-localizes again, expands into the constriction and eventually detaches from the sides to transform into a bulk mode. This transformation is consistent with the changes in the internal field landscape shown in Fig.~\ref{fig:6}(a) and the reduction of the frequency gap between the FMR and the localized mode as shown in Fig.~\ref{fig:3}(c). The observed behaviour follows from the interplay of the magnetic charges localized on the opposite sides and (top and bottom) surfaces of the constriction, and are proportional to the magnitudes of the in-plane and out-of-plane components of the magnetization, respectively. 
Both components, estimated using the mean-field approach given by Eq. (\ref{meanfield}), are shown in Fig.~\ref{fig:6}(b), where we can identify three different regimes: \emph{i}) edge localization, \emph{ii}) localization in close vicinity to the edges, and \emph{i}) bulk localization. The weakest magnetic field ($B_{0}$=0.05 T) does not saturate the sample neither out-of-plane nor in-plane. The primarily in-plane magnetization instead bends around the constriction edges to mitigate the side magnetic charges. The SW wells are located exactly at the edges but are not yet particularly deep, consistent with relatively weak mode localization seen in the top right corner of Fig.~\ref{fig:6}(b). When the field increases ($B_{0}$=0.2 T) the magnetization aligns more strongly with the in-plane component of the field, which increases the magnetic charge density at the edges, deepens the SW wells, and strengthen the mode localization, as seen at the bottom of Fig.~\ref{fig:6}(a)). This also evident from the increase of the STT efficiency shown in Fig.~\ref{fig:2} for moderate fields applied over (roughly) $\theta=60\degree$ out-of-plane, as the overlap between the edge mode and the current density increases. When the field increases further ($B_{0}$ = 0.2 - 0.5 T), the magnetization tilts more out-of-plane, in particular at the edges, which gradually redistributes the magnetic charges from the constriction edges to its surfaces. As a consequence, the SW wells detach from the edges and move gradually inwards. At yet higher fields ($B_{0} >$ 0.5 T), the surface charges dominate, the detached SW wells merge into a single shallow well closer to the constriction center, and mode localization transforms from edge to bulk.

It is now interesting to compare these linear modes with the spatial profiles of the auto-oscillations, shown in Fig.~\ref{fig:5}(b). In essentially all but a few in-plane cases, the auto-oscillations are indistinguishable from the corresponding linear localized modes. 
In stark contrast to the extended geometries where auto-oscillations emerge as either self-localized dissipative solitons or propagating SWs\cite{mohseni2013sc, bonetti2010prl}, localized constriction eigenmodes can be excited for virtually any field geometry, as their existence is not dependent on the interplay of the nonlinearity and dispersion of the system.
Only for strong in-plane, or very close to in-plane, fields, do we observe any significant difference between the linear mode and the auto-oscillation (violet box in Fig.~\ref{fig:5}(b)). As seen in Fig.~\ref{fig:3}(a), this is also accompanied by a drop in the auto-oscillation frequency, an increase in its total power, and a much larger linewidth. However, the transient behaviour of the magnetization dynamics still reveals that the auto-oscillations initially nucleate from the linear localized mode. The detailed investigation of these, likely solitonic, modes is, however, beyond the scope of the present study.

We finally want to point out that as the modes detach from the edges and move inward towards the center of the constriction, the shallow SW well allows the mode to expand quite dramatically. We can estimate the corresponding mode volume, $V$, using, \begin{equation}
V = \frac{\Delta V}{max(m_{ij})}  \sum_{i} \sum_{j} m_{ij} \nonumber
\end{equation}
where $\Delta V = \Delta x \Delta y \Delta z$ is the unit cell volume. The relative auto-oscillation volume, $V/V_{0}$, is shown in Fig.~\ref{fig:6}(b), where $V_{0}$ is obtained in the weakest in-plane field of 0.05 T. We note that $V$ increases with the out-of-plane angle of the applied field, eventually exceeding $2V_{0}$ at the highest fields. This is the underlying reason for why robust mutual synchronization of neighboring auto-oscillating constrictions can occur.\cite{Awad2016} We also observe that for small-to-moderate applied field angles (i.e.~below roughly $\theta$=40\degree), the edge mode localizes further with the applied field strength, i.e.~shrinks in volume. So direct coupling of the neighboring SHNOs should be vanishing in this case.

\begin{figure}\centering
\includegraphics[width=8.6cm]{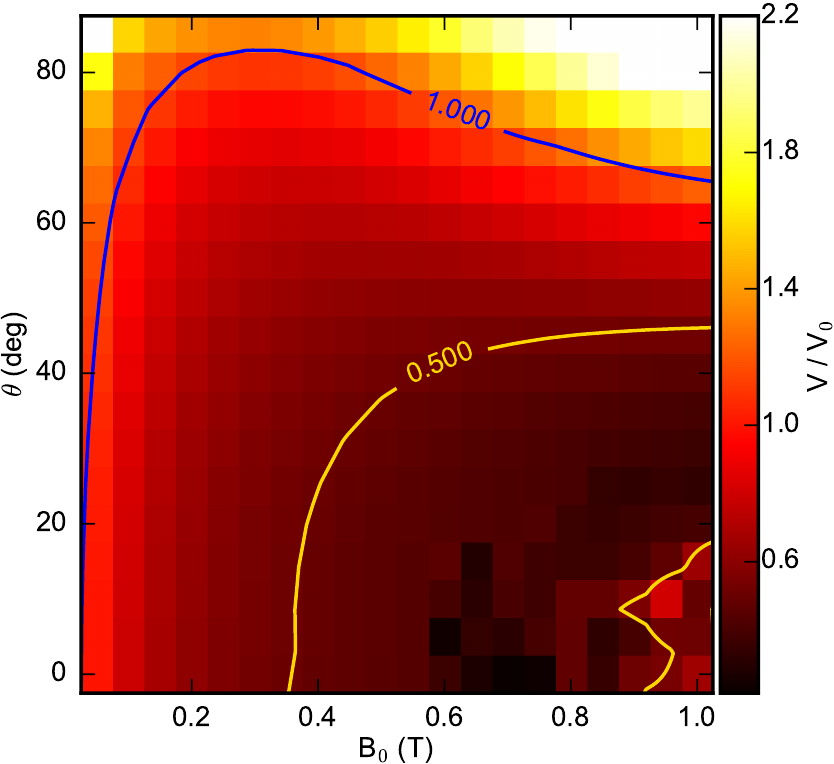} 
\caption{Relative volume of the auto-oscillating modes vs.~applied field calculated at the threshould current. }
\label{fig:7}
\end{figure}

\color{blue}

\color{black}

\section{Conclusions}
In summary, we have demonstrated, using systematic micromagnetic simulations, that auto-oscillations in constriction SHNOs originate from the linear localized eigenmodes, which appear due to the strongly non-uniform static demagnetizing field. For fields applied mostly in-plane, these modes are localized to the vicinity of the constriction edges. As the field strength and out-of-plane angle increase, the magnetic charges redistribute from the constriction sides to the surfaces, and, as a consequence, the modes change their localization character, detach from the edges, and move into the bulk of the constriction. This transformation is accompanied by a significant increase of the mode volume. Based on a macrospin model, which neglects spin wave radiation losses, we provide a qualitative description of the auto-oscillation threshold current behavior vs.~applied field. By taking into account the non-uniform character of the internal field and magnetization dynamics via a mean-field approximation, we achieve an excellent qualitative agreement with our full-scale micromagnetic simulations. In general, we observe that the stronger the localization of the edge modes, the smaller their threshold current, as \emph{i}) they experience a weaker internal magnetic field, and \emph{ii}) benefit from a higher spin current density. We find that both STT and the Oe field increase the localization of the observed modes and, correspondingly, decrease their frequencies. Furthermore, the Oe field breaks the lateral symmetry of the localized modes. We believe that our results can guide the design and implementation of interacting and mutually synchronization constriction SHNOs, emphasizing the importance of spin wave confinement for their operation.

\section{Acknowledgments}
    
    This work was supported by the Swedish Foundation for Strategic Research (SSF), the Swedish Research Council (VR), the Knut and Alice Wallenberg foundation (KAW) and the Wenner-Gren foundation. This work was also supported by the European Research Council (ERC) under the European Community's Seventh Framework Programme (FP/2007-2013)/ERC Grant 307144 ``MUSTANG''.

\section{References}
\bibliography{STO}
\end{document}